\documentclass[journal]{IEEEtran}
\usepackage{amsmath,amsfonts,amssymb}
\usepackage[colorlinks=true, allcolors=blue]{hyperref}
\usepackage{graphicx}
\usepackage{subfigure}
\usepackage{booktabs}
\usepackage{amsmath}
\usepackage{algorithm}               
\usepackage{algorithmic}            

\usepackage{cite}

\ifCLASSINFOpdf
\else
\fi
\usepackage{amsmath}
\usepackage{algorithmic}
\usepackage{fixltx2e}

\usepackage{stfloats}
\usepackage{url}

\begin{document}
%
\title{Confused Modulo Projection based Somewhat Homomorphic Encryption - Cryptosystem, Library and Applications on Secure Smart Cities}
%
%
%


\author{Xin~Jin,\IEEEmembership{}
        Hongyu~Zhang,\IEEEmembership{}
 Xiaodong~Li*,\IEEEmembership{}
Haoyang~Yu,\IEEEmembership{}
Beisheng~Liu,\IEEEmembership{}
 Shujiang~Xie,\IEEEmembership{}
Amit~Kumar~Singh,~\IEEEmembership{Member,~IEEE,}
        and~Yujie~Li\IEEEmembership{}

\thanks{Xin Jin is with Department of Cyber Security, Beijing Electronic Science and Technology Institute, Beijing 100070, China and State Key Laboratory of Cryptology, P.O. Box 5159, Beijing, 100878, China.} 

\thanks{Hongyu Zhang, Xiaodong Li and Beisheng Liu are with Department of Cyber Security, Beijing Electronic Science and Technology Institute, Beijing 100070, China. *Corresponding author, E-mail: lxdbesti@163.com.}

\thanks{Haoyang Yu is with Department of Cyber Security, Beijing Electronic Science and Technology Institute, Beijing 100070, China and Beijing University of Posts and Telecommunications, China.}

\thanks{Shujiang Xie is with School of Economics, Minzu University of China.}
\thanks{Amit Kumar Singh is with the Computer Science and Engineering Department, NIT Patna, Bihar, India.}
\thanks{Yujie Li is with Kyushu Institute of Technology 1-1 Sensui-cho Kitakyushu, 804-8550 Japan}

\thanks{Copyright (c) 2020 IEEE. Personal use of this material is permitted. However, permission to use this material for any other purposes must be obtained from the IEEE by sending a request to pubs-permissions@ieee.org.}

}

\maketitle

\begin{abstract}

With the development of cloud computing, the storage and processing of massive visual media data has gradually transferred to the cloud server. For example, if the intelligent video monitoring system cannot process a large amount of data locally, the data will be uploaded to the cloud. Therefore, how to process data in the cloud without exposing the original data has become an important research topic. We propose a single-server version of somewhat homomorphic encryption cryptosystem based on confused modulo projection theorem named CMP-SWHE, which allows the server to complete blind data processing without \emph{seeing} the effective information of user data. On the client side, the original data is encrypted by amplification, randomization, and setting confusing redundancy. Operating on the encrypted data on the server side is equivalent to operating on the original data. As an extension, we designed and implemented a blind computing scheme of accelerated version based on batch processing technology to improve efficiency. To make this algorithm easy to use, we also designed and implemented an efficient general blind computing library based on CMP-SWHE. We have applied this library to foreground extraction, optical flow tracking and object detection with satisfactory results, which are helpful for building smart cities. We also discuss how to extend the algorithm to deep learning applications. Compared with other homomorphic encryption cryptosystems and libraries, the results show that our method has obvious advantages in computing efficiency. Although our algorithm has some tiny errors ($10^{-6}$) when the data is too large, it is very efficient and practical, especially suitable for blind image and video processing.

\end{abstract}


%
\IEEEpeerreviewmaketitle

\section{Introduction}
%
%
%
%
%
%
\label{sec:intro}  
With the wide application of cloud computing technology, more and more people become used to reling on the powerful computing power and unique algorithm of cloud to process their image and video information. Although image and video intelligent cloud service brings great convenience to users, it also give rise to the security issues of users' privacy information. Once the attacker successfully attacks the cloud storage server, the attacker misuses these image data carrying a large amount of user privacy for illegal behavior. 

Blind computing is a method and technology for processing the original content of the data without touching it. Visual blind computing is a kind of blind computing. It includes not only simple visual data processing such as target detection and target recognition, but also can be applied to various complex visual data processing, such as semantic segmentation~\cite{Quan2019An,Zhou2019Multi}. As early as 2006, Avidan et al. \cite{avidan2006blind} proposed the concept of blind vision and a blind computing version of the classic Viola \& Jones target detection algorithm using standard cryptographic tools. The server performs remote facial blind detection without leaking the parameters of the target detector. But the effectiveness of cryptographic algorithm is low, it takes several hours to scan an ordinary image. 
In 2009, Upmanyu et al. proposed a framework~\cite{upmanyu2009efficient} to carry out privacy-preserving surveillance based on secure multi-party computation \cite{rabin1989verifiable}. It splits each frame into a set of random images and sends them to different servers. Each image by itself does not convey any meaningful information about the original frame, while collectively, it retains all the information. This solution is derived from a secret sharing scheme based on the Chinese Remainder Theorem (CRT) \cite{ore1988number}, suitably adapted to image data. But the method leads to satisfactory results only in controlled environments. It can become problematic in practical applications\cite{jin2016ppvibe}. The conditions of use are harsh. This solution requires multiple servers to work at the same time to get the results, and if the data of multiple servers is leaked, the attacker can recover the data to get the original data.
In 2010, Osadchy et al. proposed a privacy-protected face recognition protocol~\cite{osadchy2010scifi}. This protocol realizes face recognition with privacy protection by computing Hamming distance between compact representation of face image and privacy protection.
In 2013, Chu et al. proposed a real-time detection method for video moving targets with privacy protection in a cloud environment~\cite{10.1145/2502081.2502157}. In this method video frames are encrypted by color flipping, pixel rearrangement, matrix multiplication, and quantization. This process does not change the Gaussian statistical properties of the video frame, so the server can still get the moving object detection results of encrypted frames through the video moving object detection method based on Gaussian mixture model.
In 2014, Chu et al. proposed a privacy preserving multi camera target tracking method~\cite{10.1145/2647868.2655010}. The distance calculation method based on garbled circuit arranges more computation to a single monitoring terminal, which reduces the cooperative calculation of multiple monitoring terminals. Under the condition that the accuracy rate of target tracking is approximate or higher than that of traditional non privacy preserving tracking algorithm, the overall time cost is only a few seconds higher than that of non-privacy preserving tracking algorithm.
In 2015, Bost et al. proposed a machine learning classification protocol for encrypted domain data, which can be applied to face detection of user video images~\cite{Bost2015Machine}. In this work, three privacy preserving classifiers are constructed: hyperplane decision, naive Bayes and decision tree, which can be combined by AdaBoost.
In 2017, Jin et al. proposed a privacy preserving face image detection protocol based on random base image representation in typical cloud computing environments, and implemented a privacy preserving face image detection algorithm on OpenCV~\cite{Xin2017Efficient}.
In recent years, with the rise of deep learning research, visual blind computing for deep learning has become a new research direction.

In order to improve the performance of blind computing, based on multi-party computing, homomorphic cryptography, functional encryption, security sharing and other types of cryptography method are continuously applied to blind calculations. The research of secure multi-party computing is mainly about how to calculate a contract function safely without a trusted third party, and ensure that the input of each participant is secret. Secure multiparty computing was first proposed by Yao~\cite{yao1982protocols} in 1982. In recent years, there have been some new research developments in secure multi-party computing ~\cite{Hirt2013A,Lee2014,Chuan2019Secure,Eunkyung2020Towards}. The operating efficiency of secure multi-party computing is still significantly lower than that of computing using a trusted third party. For this reason, using other cryptographic or mathematical tools to construct secure multi-party computing protocols for specific problems is also a research direction. 
Functional encryption was first proposed by Dan et al.~\cite{boneh2011functional} in 2011, that is, under the premise of authorized key, the specific function of plaintext can be calculated through ciphertext, but plaintext cannot be obtained. In recent years, function encryption has made some theoretical progress, although there is still a big gap in efficiency from the practical~\cite{goldwasser2013reusable,Junichi2020Unbounded}.

The concept of homomorphic encryption was first proposed in 1978 \cite{rivest1978data}. The emergence of homomorphic encryption technology allows users to make a meaningful calculation on cipher text directly, and the result is still a cipher text. At present, homomorphic encryption has no unified classification standard, but it can be divided into partial homomorphic encryption (PHE), somewhat homomorphic encryption (SWHE) and full homomorphic encryption (FHE) according to its development stage, types and times of ciphertext operation. PHE only supports a single type of ciphertext field homomorphism (additive homomorphic properties or multiplicative homomorphic properties). Paillier encryption scheme~\cite{paillier1999public} proposed in 1999 is the most widely used PHE scheme. It satisfies the additive homomorphic properties and scalar multiplicative homomorphic properties. Partial homomorphic encryption schemes are relatively mature in application due to their simple structure and high efficiency. However, its application scope is still limited because it only supports a single type of homomorphic operation in ciphertext field. SWHE scheme can support both additive homomorphic properties and multiplicative homomorphic properties, but it can only do limited times of addition or multiplication operations. In 2005, Boneh et al. Proposed a BGN scheme~\cite{boneh2005evaluating} with semantic security, which is constructed by using the structural characteristics of algebraic rings. This is the first scheme that can support both unlimited times of ciphertext addition and one-time multiplicative homomorphism operation. FHE scheme can realize the additive homomorphic properties and multiplicative homomorphic properties, and can do any times of addition and multiplication operations. In 2009, Gentry constructed a FHE scheme~\cite{gentry2009fully1} based on ideal lattice for the first time. A SWHE scheme is constructed firstly in this scheme which can only evaluate low depth circuits. To ensure security, "noise" is added to the ciphertext. However, this noise will increase with each operation until it leads to decryption error. Therefore, it is necessary to simplify the decryption circuit so that it can homomorphically calculate its own decryption circuit. Finally, SWHE can be transformed into FHE scheme which can homomorphically calculate any circuit by bootstrap.
Since then, many scholars have optimized and simplified Gentry's scheme. In 2012, Brackerski et al. manage the noise level by modulus switching and construct a leveled fully homomorphic encryption schemes (capable of evaluating arbitrary polynomial-size circuits), without Gentry's bootstrapping procedure(BGV)~\cite{Brakerski2012}.In the following years, many scholars have improved the BGV scheme, in which Halevi et al. implemented BGV scheme with bootstrapping~\cite{halevi2015bootstrapping} and then greatly improved the efficiency of the scheme~\cite{halevi2018faster}. In 2012, Fan et al. ported Brakerski's FHE scheme(BFV) from the LWE to the RLWE setting and provided a detailed analysis of all subroutines involved such as multiplication, relinearisation and bootstrapping~\cite{fan2012somewhat}. In 2016, Bajard et al. suggested a way to entirely eliminate the need for multi-precision arithmetic, and presents techniques to enable a full RNS implementation of FV-like schemes~\cite{bajard2016full}. In 2016, Cheon et al. proposed a method to construct an approximate homomorphic encryption scheme (CKKS)~\cite{cheon2017homomorphic}. It supports floating-point homomorphism operation, and can approximate add and multiply encrypted messages. A follow up work in 2018 by Cheon et al. proposed a method to refresh ciphertexts of CKKS scheme and make it bootstrappable for the evaluation of an arbitrary circuit~\cite{cheon2018bootstrapping}. Soon after, Chen et al. improved the bootstrapping result in Cheon's work by replacing the Taylor approximation of the sine function with a more accurate and numerically stable Chebyshev approximation and designing a modified version of the Paterson-Stockmeyer algorithm for fast evaluation of Chebyshev polynomials over encrypted data~\cite{chen2019improved}. In recent years, some homomorphic encryption libraries based on these schemes have also been developed, such as Microsoft SEAL Library (support both BFV scheme and CKKS scheme)~\cite{sealcrypto}, IBM helib Library (support both BGV scheme and CKKS scheme), etc. At present, the performance of fully homomorphic encryption is still far from practical application, and partial homomorphic encryption combined with specific visual application scenarios is more widely used in practical.

In this paper, we propose a single-server version of blind processing method--CMP-SWHE: Confused Modulo Projection based Somewhat Homomorphic Encryption algorithm. CMP-SWHE algorithm is based on the modulo projection theorem, the equivalence properties of congruences and the Chinese Remainder Theorem (CRT). This algorithm belongs to somewhat homomorphic encryption algorithm. On the client side, the original data is encrypted by amplification, randomization, and setting confusing redundancy. Operating on the ciphertext on the server side is equivalent to operating on the original data. As an extension, we designed and implemented a blind computing scheme of accelerated version based on batch processing technology to improve efficiency. We did a lot of experiments to test our scheme and applied it to blind foreground extraction, blind optical flow tracking and blind face detection. We also extend the algorithm to applications of deep learning. Experimental results show that although our algorithm has some tiny errors ($10^{-6}$), it is very efficient and practical, especially suitable for image and video processing, which does not require high accuracy.

The main contributions of this paper: 

1. We propose a single-server version of Confused Modulo Projection based Somewhat Homomorphic Encryption algorithm (CMP-SWHE), which is based on the modulo projection theorem, the equivalence properties of congruences and the Chinese Remainder Theorem (CRT); 

2. We design and implement a blind computing scheme of accelerated version based on batch processing technology to improve efficiency;

3. We design and implement a practical general library based on this CMP-SWHE algorithm, and experiments show that our library has great efficiency advantages in application; 

4. We apply our scheme to various blind vision problems such as blind foreground extraction, blind optical flow tracking and blind object detection;

5. We discuss how to extend the algorithm to applications of deep learning.

\section{CMP-SWHE Cryptosystem}
\subsection{Concept introduction}
\subsubsection{Modulo Projection}
\begin{figure}[htbp]
\centering
\includegraphics[width=\linewidth]{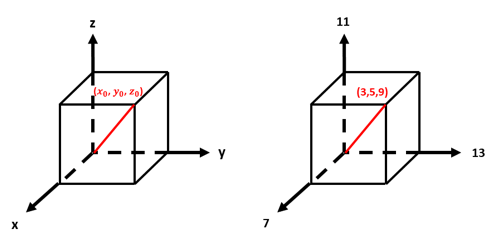}
\caption{Schematic diagram of modulo projection}
\label{fig:1}
\end{figure}

In geometry, any point in n-dimensional space has a projection on each coordinate axis, and the projection on three axes can also determine the unique point in space.
The modulo operation can be analogized here. The $n$ two mutually prime modulo bases can be regarded as n coordinate axes, and an integer can be regarded as a point in the coordinate system. The result of modulo operation of this integer on a modulo base can be considered as a modulo projection on the coordinate axis. As shown in the figure \ref{fig:1}, 31 is a point on the three-dimensional coordinate axis (7,11,13). Its projection on the "7 axis" is 3, the projection on the "11 axis" is 9, and the projection on the "13 axis" is 5. We can also call these three axes as modulo bases.

\subsubsection{Notation Table}
To improve the understanding of the paper, we list all the notations and their meanings in Table~\ref{tab:notations}.

\begin{table}[!t]
\caption{Notation Table}
\label{tab:notations}
\centering
\begin{tabular}{|c|c|}
\hline
Notation & Meaning    \\ 
\hline
$a$ & Amplification factor\\ 
\hline
$\eta$ & Random number\\ 
\hline
$B$ & Modulo bases\\ 
\hline
$b_{i}$ & The i-th module base\\ 
\hline
$S$ & Position template\\ 
\hline
$s_{i}$ & Correct modulo projection index of the i-th module base\\ 
\hline
$P$ & Original data\\ 
\hline
$p$ & Correct modulo projections\\ 
\hline
$r$ & Confusing redundant modulo projections\\ 
\hline
$e$ & Encrypted original data\\ 
\hline
\end{tabular}
\end{table}

\subsection{Algorithm}
\label{mod}
CMP-SWHE algorithm consists of four parts: generating public key and private key, encryption, blind computing, decryption. In the first part, public key and private key will be generated at the client. In the second part, the original data will be encrypted at the client by amplifying, randomizing, calculating modulo projection, and setting confusing redundant modulo projections. In the third part, the server performs addition, subtraction, multiplication and other operations on the encrypted data, which is equivalent to the operation on the original data. In the last part, the client takes out the correct model projection according to the key, and decrypts the results according to CRT.
\subsubsection{Generation of public key and private key}
\label{221key}
In CMP-SWHE, the public-private key mechanism is used to protect privacy. The former includes amplification factor ($a$) and modulo bases ($B$). The latter includes these two keys and position template ($S$), which contains the correct position of modulo projection of each modulo base. The maximum value of the random number $\eta$ is also set in the private key. Every user can use their unique user key ($U$) to generate public and private key on the client side. Users keep their private keys, while the server only has public keys. In this way, we can ensure that only the user can get the correct answer and others cannot get original data through the server. The specific process is shown in Algorithm~\ref{221}.

Since both modulo base group $B$ and position template $S$ are generated by the user key $U$, the length of the user key $U$ determines the maximum amount of redundancy and the maximum amount of modulo bases.

%

\begin{algorithm}[htb]         
\caption{Generation of public key and private key}             
\label{221}                  
\begin{algorithmic}[1]                
\REQUIRE ~~\\  User Key ($U$)                   
    
\ENSURE ~~\\   Public key and private key

\STATE We note that the time stamp when the client applies for this service is $T$.

\STATE Select $n$ prime modulo bases to form a modulo base pool $B_{n}=\left\{b_{1}, b_{2}, \cdots, b_{n}\right\}$.$b_{i}$ is prime number.

\STATE Define the number of modulo bases $N (n> N)$ and the number of confusing redundant modulo projections $M-1$, that is, there are $M$ modulo projections in total.

\STATE Encrypt the timestamp $T$ by AES (ECB mode) with $U$ as the key to obtain a ciphertext ($C_{1}$). After $C_{1}$ is converted to binary, every $6$ bits corresponds to an index ($s_{i}$) for position of correct modulo projections. We make up the position template $S$ with $N$ position indexes. $S=\left\{s_{1}, s_{2}, \cdots, s_{N}\right\}, \quad 0 \leq s_{i} \leq M-1, \quad i=1,2, \ldots, N$.

\STATE Encrypt ciphertext $C_{1}$ by AES (ECB mode) with $U$ as the key to obtain a ciphertext ($C_{2}$). After $C_{2}$ is converted to binary, every $6$ bits corresponds to an index ($b_{i}$) for number of modulo base. We make up the modulo base group $B$ with $N$ modulo base indexes. The elements $b_{1}, b_{2}, \cdots, b_{N}$ corresponds to the index in modulo base pool $B_{n}$. $B=\left\{b_{1}, b_{2}, \cdots, b_{N}\right\}, i=1,2, \ldots, N$.
             
\end{algorithmic} 
\end{algorithm} 

\subsubsection{Encryption}
\label{222enc}
~\\
In section \ref{221key}, we get public key and private key. Now, we can use these keys to encrypt the original data on the client side. Suppose there are $20$ modulo bases and $63$ confused redundancies. First, the original data modulo each corresponding modulo base, and get $20$ modulo projections. Then, for each modulo base, we add $63$ confusing redundant modulo projections. After these two operations, we will get a $20 \times 64$ ciphertext. For each modulo base, the location of the correct modulo projection in these $64$ projections is determined by the location template ($S$) in the private key.

In reality, not only the client has the privacy to be protected, but also the server. For example, the training model, parameters or algorithms are provided by the server as business secrets, and the server does not want them to be disclosed. However, the server can't hold the private key in CMP-SWHE, so that the server can't insert the real modulo projection into confused redundant modulo projections according to the private key like the client. To solve this problem, we provide different encryption interfaces for the client side and the server side. The client encrypts according to the private key, and the server encrypts according to the public key. When encrypting according to the public key, redundant projections should also be set, but they are only for the unity of the cipher structure, rather than to confuse the real modulo projection. It is not necessary to protect the location of the real modulo projection, so we store the real modulo projection in each location ($64$).

Take encrypting original data $P$ to ciphertext $e$ as an example. The specific process is shown in Algorithm~\ref{222}.

%

\begin{algorithm}[htb]         
\caption{Encryption}             
\label{222}                  
\begin{algorithmic}[1]                
\REQUIRE ~~\\  Original data ($P$)                  
    
\ENSURE ~~\\   Encrypted original data (Ciphertext $C$)                   

\STATE Generate a random integer array $R=\left\{R_{1}, R_{2}, \ldots, R_{M}\right\}$ as confusing redundant data.

\STATE Scale up and randomize original data $P$.
\begin{equation}
\label{eq:amp}
P^{\prime}=a P+\eta
\end{equation}

\STATE Scale up and randomize confusing redundant data $R$. This step is not required for encryption based on public key on the server side. 

$$\begin{array}{c}
R_{i}^{\prime}=a R_{i}+\eta, R^{\prime}=\left\{R_{1}^{\prime}, R_{2}^{\prime}, \ldots, R_{M-1}^{\prime}\right\} \\
i=1,2, \ldots, M-1
\end{array}$$

\STATE Calculate the correct modulo projections $p$.
$$
p=\left\{p_{1}, p_{2}, \cdots, p_{N}\right\}, p_{i}=P^{\prime} \bmod b_{i}, \quad i=1,2, \ldots, N
$$

\STATE Calculate confusing redundant modulo projections $r$.

(1) For encryption based on private key on the client side:

$$\begin{array}{c}
r_{i}=\left\{r_{i_{1}}, r_{i_{2}}, \ldots, r_{i_{M}}\right\}, i=1,2, \ldots, N \\
r_{i_{j}}=R_{i}^{\prime} \bmod b_{i}, j=1,2, \cdots, M \\
r=\left\{r_{1}, r_{2}, \cdots, r_{N}\right\}
\end{array}$$

(2) For encryption based on public key on the server side:

$$\begin{array}{c}
r_{i}=\left\{r_{i_{1}}, r_{i_{2}}, \ldots, r_{i_{M}}\right\}, \quad i=1,2, \ldots, N \\
r_{i_{j}}=P^{\prime} \bmod b_{i}, \quad j=1,2, \cdots, M \\
r=\left\{r_{1}, r_{2}, \quad \cdots, \quad r_{N}\right\}
\end{array}$$

\STATE Insert the correct modulo projections $p$ into confusing redundant modulo projections $r$ according to position template $S$ and then obtain the $M \times N$ ciphertext $e$.

(1) For encryption based on private key on the client side:



$$\begin{aligned}
&e_{i j}=\operatorname{encrypt}(P)=\left\{\begin{array}{l}
r_{i j}, j \neq S_{i} \\
p_{i}, j=S_{i}
\end{array}\right.\\
&i=1,2, \dots, N, \quad j=1,2, \dots, M
\end{aligned}$$

(2) For encryption based on public key on the server side:

$$e_{i_{j}}=\text {encrypt}(P)=p_{i}, j=S_{i}, i=1,2, \ldots, N$$
      
\end{algorithmic} 
\end{algorithm} 

\subsubsection{Blind Operation}
\label{223op}
~\\
In this part, we will add, subtract or multiply the ciphertext obtained in section~\ref{222enc} on the server side. Because of the homomorphism of modulo operation, the operation of ciphertext is equivalent to the operation of plaintext. At the end of this part, we will get the blind calculation results of the ciphertext. In this process, the server does not have the location template ($S$) in the private key, it will not know  which are correct modulo projections. The specific process is shown in Algorithm~\ref{223}.

%

\begin{algorithm}[htb]         
\caption{Blind Operation}             
\label{223}                  
\begin{algorithmic}[1]                
\REQUIRE ~~\\  Encrypted original data (Ciphertext $e$)                
    
\ENSURE ~~\\   Result of $e$ (f($e$))                      

\STATE Make the expression homogeneous by multiply items with low order by amplification $a$. 
The concept of order is described in detail in section~\ref{order}. 

\STATE According to Congruent equivalence property and its extension theorem bellow, we can add, subtract, multiply and power (Collectively referred to as $f$) ciphertext at the server.
$$\begin{array}{c}
(x+y) \bmod b=((x \bmod b)+(y \bmod b)) \bmod b \\
(x-y) \bmod b=((x \bmod b)-(y \bmod b)+b) \bmod b \\
(x * y) \bmod b=((x \bmod b) *(y \bmod b)) \bmod b \\
x^{y} \bmod b=(x \bmod b)^{y} \bmod b
\end{array}$$           
Suppose that ciphertext $x, y, z$ are obtained by encrypting plaintext $X, Y, Z$.
$x$ = encrypt ($X$), $y$ = encrypt ($Y$), $z$ = encrypt($Z$).
A series of blind operations are performed on these three ciphertexts to obtain a result $f(x, y, z)$ of size $M \times N$. By decrypting it, we can get the calculation result of the original data $x, y, z$. $f(X, Y, Z) = decrypt(f(x, y, z)) $
\end{algorithmic} 
\end{algorithm} 

\subsubsection{Decryption}
\label{224dec}
~\\
In this part, we will decrypt the blind calculation results of ciphertext obtained in section \ref{223op} on the server side. First, extract blind operation results of correct modulo projections ($20$) from blind operation results of all the module projections ($20 \times 64$). Then, the blind calculation results of the original data will be obtained through the Chinese Remainder Theorem. The specific process is shown in Algorithm~\ref{224}.

%
%

\begin{algorithm}[htb]         
\caption{Decryption}             
\label{224}                  
\begin{algorithmic}[1]                
\REQUIRE ~~\\   Result of $e$ (f($e$))                     
    
\ENSURE ~~\\    Result of $P$ (f($P$))                     

\STATE Takes out the correct modulo projections of calculation results $p=\left\{p_{1}, p_{2}, \cdots, p_{N}\right\}$ according to the position template $S$.

\STATE Decrypt the results of original data $f(X, Y, Z)$ according to CRT.

$$f(X, Y, Z)=\frac{\left(\sum_{i=1}^{i=N} p_{i} B_{i} B_{i}^{-1}\right) \bmod B_{s}}{a^{t}}$$

where $t$ is the order of the polynomial, $B_{s}=\prod_{i=1}^{i=N} b_{i}$, $B_{i}=B_{s} / b_{i}$, $B_{i} B_{i}^{-1}=1 \bmod b_{i}$.                 
\end{algorithmic} 
\end{algorithm} 

\subsubsection{Parameter setting}
\label{parametersetting}
~\\
1. Every modulo base $b_{i}$ in modulo base group is prime number. This is the requirement of Chinese Remainder Theorem.

2. The random number $\eta$ corresponding to each original data is different. In addition, $\eta_{max}$$>\max \left(b_{i}\right)$, so that original data can be cover up after modulo operation. The reason for this setting is discussed in detail in section~\ref{securityanalysis}.

3. $a >>$$\eta$. Only in this way can the random number and its derived noise be eliminated after CRT. The reason for this setting is discussed in detail in section~\ref{errora}. 

4. $w *(a * \max (P))^{t}<\prod_{i=1}^{i=N} b_{i}$, where $w$ is the number of terms of the polynomial and $t$ is the order of the polynomial.

\subsubsection{Examples}
\label{examples}
~\\

\begin{figure}[htbp]
  \centering
  \includegraphics[width=\linewidth]{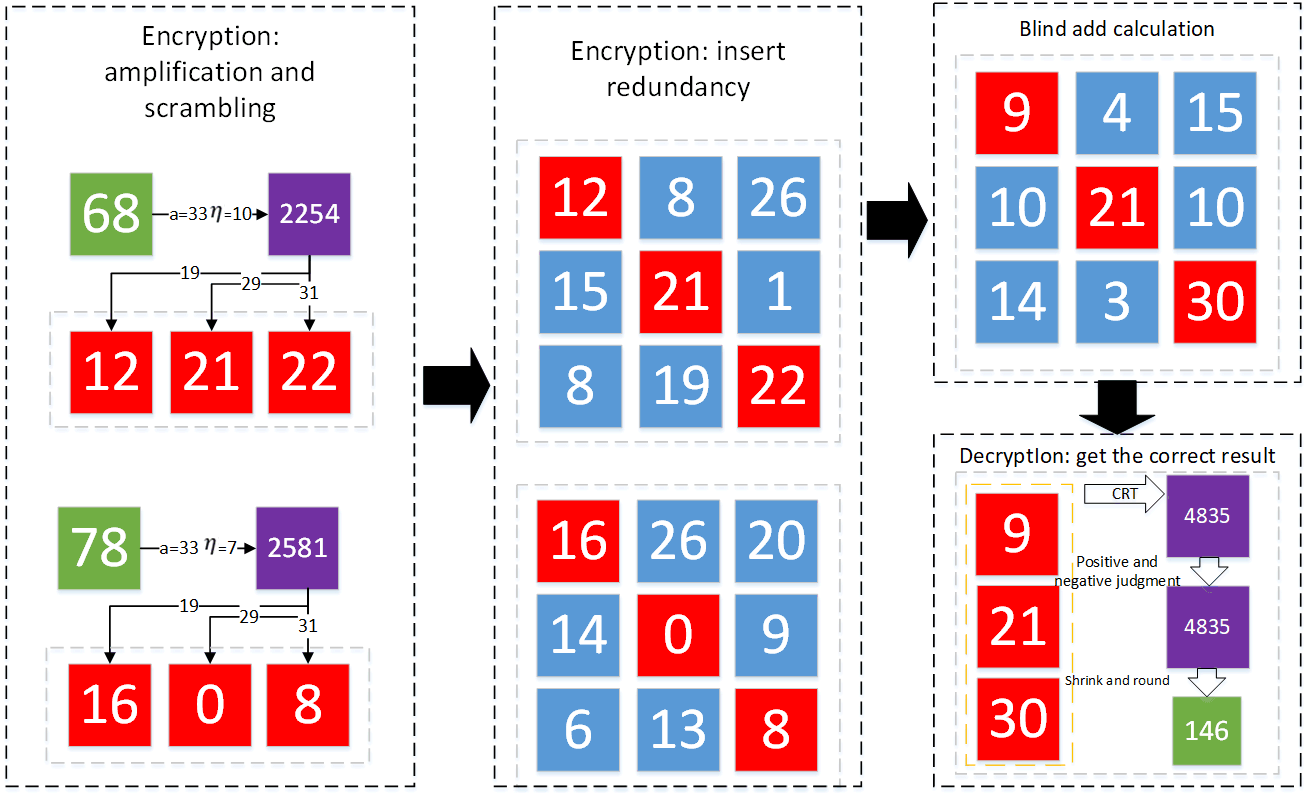}

\caption{Example of blind addition of two numbers}
\label{fig:exa}       
\end{figure}

In Figure \ref{fig:exa}, We give an example to describe the whole process of blind adding two pixels in detail. Consider these two pixels with value $x=68$ and $y=78$. Let the amplification factor $a$ be $33$, the number of modulo bases $N$ be $3$, the modulo base $B=\left\{b_{1}, b_{2}, b_{3}\right\}$, where $b_{1}=19$, $b_{2}=29$,$b_{3}=31$ and the number of redundancies $M$ be $2$. The parameters set here are just for the convenience of display. In the actual process, the parameters will be different.

In the encryption stage, suppose the random numbers are $10$ and $7$, then the resulting of Equation~\ref{eq:amp} is $68\times33+10=2254$ and $78\times33+7=2581$. 
After encryption, the plain data $x$ becomes $\operatorname{encrypt}(x)=\{[12,8,26],[15,21,1],[8,19,22]\}$, and the plain data $y$ becomes $\operatorname{encrypt}(y)=\{[16,26,20],[14,0,9],[6,13,8]\}$, the client sends $\operatorname{encrypt}(x)$ and $\operatorname{encrypt}(y)$ to the server.

In the blind calculation stage, the server performs addition operation on the encrypted data, and performs modulo operation according to the modulo base $B$, and finally obtains the calculation result as $\operatorname{encrypt}(x+y)=\{[9,4,15],[10,21,10],[14,3,30]\}$, and sends the calculation result to the client.

In the decryption stage, the correct modulo projections of calculation results $d=\left\{d_{1}, d_{2}, d_{3}\right\}=\left\{9,21,30\right\}$. Using the Chinese remainder theorem, we get $4835$. Then, we reduce the result by $33$ times and get $146.515152$. Finally, we remove the decimal and keep the integer part, and get $146$($68+78=146$). The reason why there is a decimal is that we add random numbers to cover the real data during encryption.

\subsection{Analysis of Privacy}
\label{securityanalysis}

\begin{figure}[htbp]
\centering
\includegraphics[width=\linewidth]{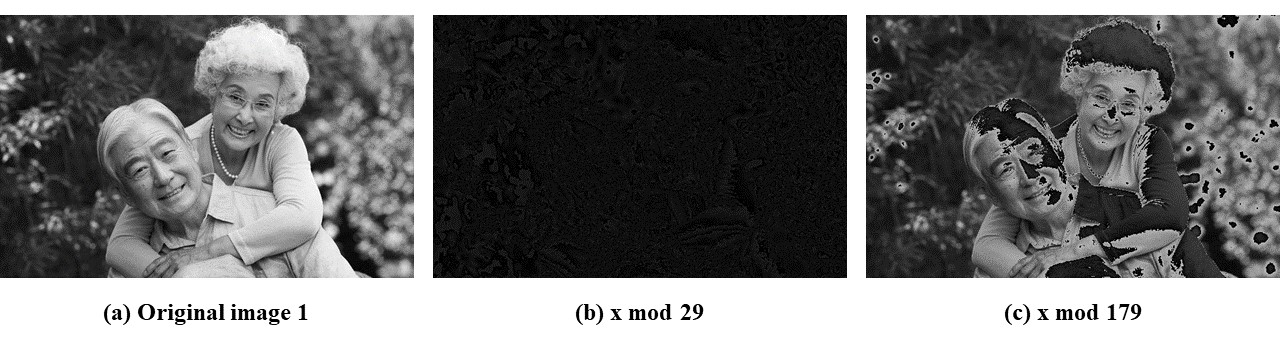}
\caption{The different effect of the size of modulo base on encryption}
\label{fig:modbase}       
\end{figure}

\begin{figure}[htbp]
\centering
\includegraphics[width=\linewidth]{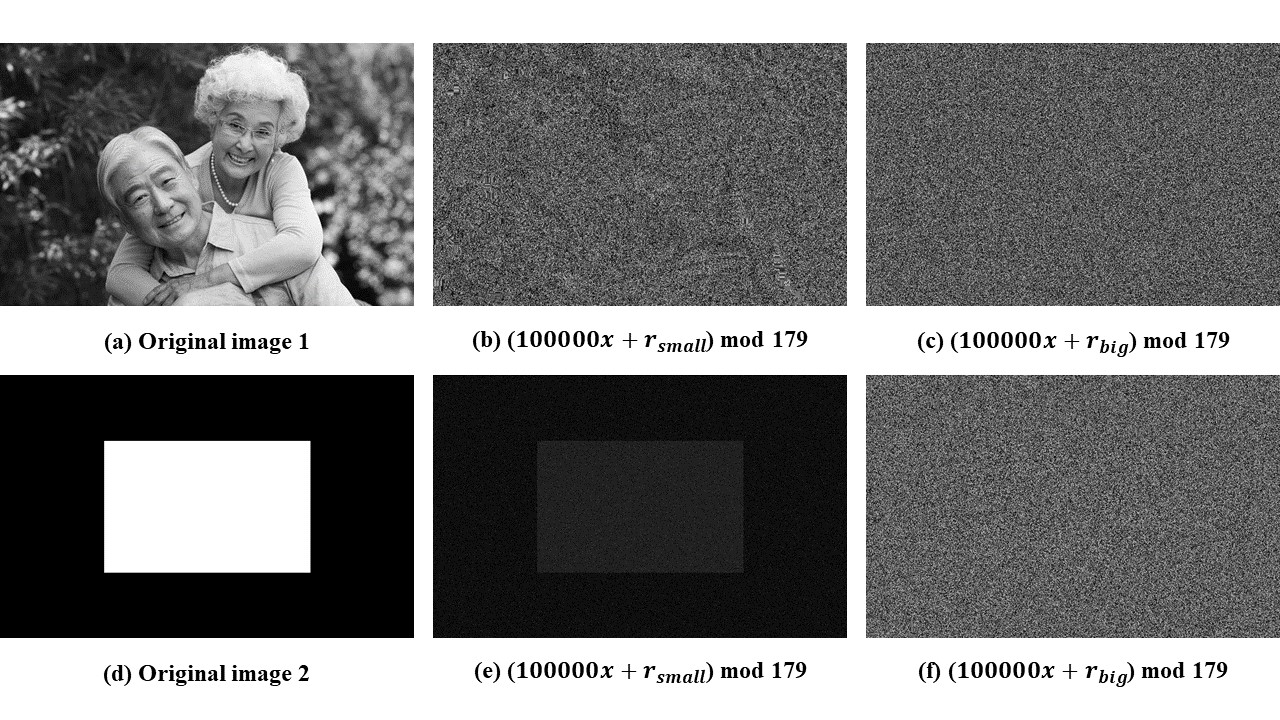}
\caption{The different effect of range of random number on encryption}
\label{fig:random}       
\end{figure}

The original data protection mechanism of first enlarging the original data by a certain multiple $a$ and then adding a small random number $\eta$ (Formula~\ref{eq:amp}) has been used in Upmanyu's work \cite{upmanyu2009efficient}. Here is a brief introduction to its principle: As can be seen from Figure~\ref{fig:modbase}, if we only do modulo operation on the original data, when the modulo base is not small enough, due to the values of adjacent pixels are correlated, the server can still see the general appearance of the original image. Therefore, we add a different random number to each pixel in the original image to achieve the purpose of scrambling. But with different random numbers, it is difficult to recover the original data after decryption. Therefore, we first multiply the original data by a huge amplification $a$, and then add a small random number. As can be seen from Figure~\ref{fig:random}(c), after this processing, the server cannot distinguish the original image. After decryption, we divide the original data by this amplification $a$. The error caused by the added random number can be ignored after dividing by the huge amplification $a$. However, if there are too many times of multiplication, or the multiplier is too large, this error will accumulate to a point that cannot be ignored. In section \ref{errora}, we analyze the factors that influence this error.

%
%

\begin{figure}[htbp]
  \centering
  \includegraphics[width=\linewidth]{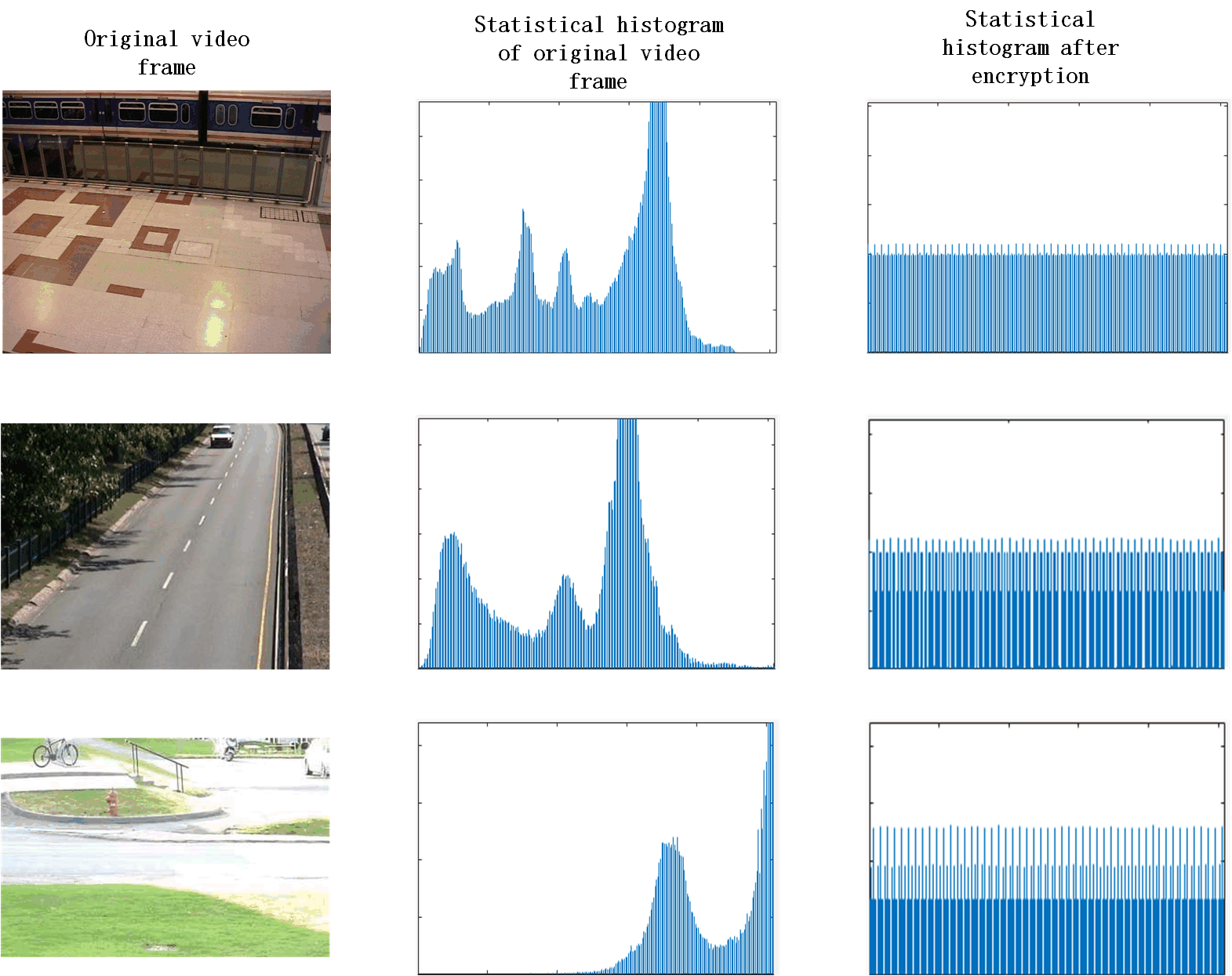}

\caption{Gray histogram comparison between plain video frames and encrypted video frames.}
\label{fig:sta}       
\end{figure}

We now analytically show that it is difficult to make statistical attacks on the frames after amplification and randomization under an optimum parameters selection. If the random number $\eta$ follows the uniform distribution, $U(0, r_{max}-1)$, the distribution of $\eta^{\prime}=\eta \% b_{i}$ would be:  
$$\operatorname{Pr}\left(\eta^{\prime}=x\right)=\left\{\begin{array}{l}\frac{l+1}{\eta_{\max }}, x<\eta_{\max } \% b_{i} \\ \frac{l}{\eta_{\max }}, x \geq \eta_{\max } \% b_{i}\end{array}\right.$$
, where $x \in\left[0, b_{i}\right), l=\left\lfloor\frac{\eta_{max}}{b_{i}}\right\rfloor$. $\eta_{max}$ is set in the private key and known only to the client.
In the ideal state, if $\eta_{max}$ is the common multiple of each modulo base $b_{i}$, then $\eta^{\prime}$ completely follows the uniform distribution. That is, after the frame is enlarged and randomized, pixel values will be uniformly distributed and information in the original frame will not be revealed. But if $\eta_{max}$ is not multiple of $b_{i}$, for the modulo projection $x^{\prime}$ of pixel $x$, which is amplified and randomized, $\operatorname{Pr}\left(x \in\left[0, \eta_{\max } \% b_{i}\right)\right)>\operatorname{Pr}\left(x \in\left[\eta_{\max } \% b_{i}, b_{i}\right)\right)$. That is, $x^{\prime}$ is slightly more likely to be in a particular region. Although the probability difference is only $\frac{1}{\eta \max }$, in some extreme cases, hackers may find the rule.
Figure~\ref{fig:random} shows the effect of random numbers of different sizes on encryption. In \ref{fig:random}(b) and \ref{fig:random}(e),we set $\eta_{max}<20$ ,while in \ref{fig:random}(c) and \ref{fig:random}(f) $\eta_{max}>\max \left(b_{i}\right)$. Perhaps in some normal images, like Figure~\ref{fig:random}(a), the random number size has little effect on the encryption results. But in some extreme cases, such as in Figure~\ref{fig:random}(d), if the random number is too small, we can still see the black-and-white division line of the original image in the ciphertext image \ref{fig:random}(e).Therefore, on the basis of Upmanyu's paper~\cite{upmanyu2009efficient}, we have made a more strict limit on the range of random numbers as $U\left(0, \eta_{\max }\right)$, $\max \left(b_{i}\right)<\eta_{max}<<a$ in Section~\ref{parametersetting}. Further, $\eta_{max}$ should be as large as possible in this range. Because the larger the $\eta_{max}$, the larger the $l$ and the smaller the $\frac{1}{\eta \max }$. When the $\frac{1}{\eta \max }$ is small enough, the distribution can be regarded as approximately uniform. In Figure~\ref{fig:sta}, we show the gray histogram of the original frame and the encrypted frame, which can reflect the relationship between the frequency of gray pixels and the gray level. It can be seen that the encrypted histogram is flat, which makes hackers unable to attack by analyzing its statistical characteristics. In other words, our scheme has a strong ability to resist statistical attacks.

Upmanyu's work~\cite{upmanyu2009efficient} split each frame after amplification and randomization into a set of random images and send to several servers to ensure the information security. In our scheme, we send this set of random images to the same server for processing. So, how can we prevent privacy information leakage that may result from this change? We add a certain number of confusing redundant modulo projections for each modulo base after calculating the correct modulo projection. Suppose we have $N$ modulo bases and $M-1$ confusing redundancies. Because the server does not hold private key, the probability of getting $N$ correct modulo projections without location template is $\frac{1}{M^{N}}$. When $M$ and $N$ are larger, the attack is more difficult and the security is higher. Ideally, when $M$ and $N$ are set to be large enough, the attacker will not be able to carry out exhaustive attacks. However, if $M$ and $N$ are very large, the computing cost will also be large. Therefore, users need to find a balance between security and efficiency according to their own needs. How the modulo bases and confusing redundancies influence the efficiency is analyzed in detail in Section~\ref{performance}.


\begin{figure}[htbp]
\centering
\includegraphics[width=\linewidth]{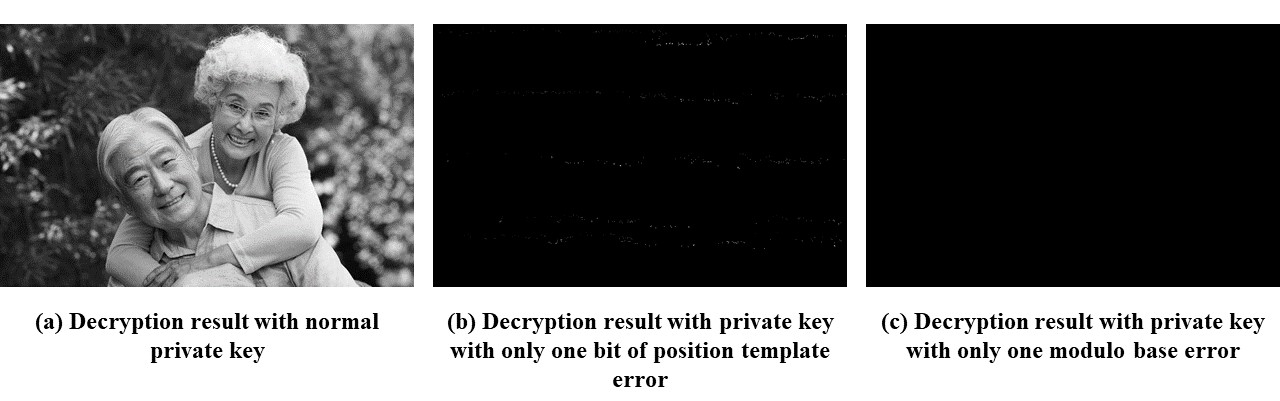}
\caption{Key sensitivity.}
\label{fig:sensitivity}   
\end{figure}

In order to verify the sensitivity of the key more intuitively, we use the correct private key to encrypt Figure~\ref{fig:random}(a), and then use the wrong private key to decrypt. The decryption result is shown in Figure~\ref{fig:sensitivity}. In Figure~\ref{fig:sensitivity}(a), we change only one module base in the private key, and in Figure~\ref{fig:sensitivity}(b), we change only one module base's location template. It can be seen that although the slight change of the private key has a huge impact on the decryption result, which can prove that our scheme is safe.

In conclusion, based on the idea of frame splitting proposed in Upmanyu's work~\cite{upmanyu2009efficient}, we changed the multi-server version into a single-server version. In order to avoid the security problems caused by this change, we add the confused redundancy mechanism and the corresponding public-private key mechanism. To further prevent leakage of privacy information in some extreme cases, we also standardize the range of random numbers. In addition, in order to prove that our single server version is consistent with the function in Upmanyu's work~\cite{upmanyu2009efficient}, we reproduce all its applications in Section~\ref{app}.

\subsection{Error Analysis}
\label{errora}
~\\
There is a limitation in CMP-SWHE algorithm, that is, random numbers are added in encryption, which cannot be eliminated without decryption.

For example, if we want to add plaintext $X$ and $Y$, we first encrypt the plaintext to get ciphertext $\left(a X+\eta_{1}\right) \bmod B$ and $\left(a Y+\eta_{2}\right) \bmod B$, where $a$ is the amplification factor, $\eta_{1}$ and $\eta_{2}$ are random numbers, and $B$ is the modulo group. The result of blind addition is $\left(a X+\eta_{1}+a Y+\eta_{2}\right) \bmod B$. After decryption, we get the addition result of plaintext $X$ and $Y$: $\frac{a X+\eta_{1}+a Y+\eta_{2}}{a}=X+Y+\frac{\eta_{1}+\eta_{2}}{a}$. This result has an additional error term $\frac{\eta_{1}+\eta_{2}}{a}$ than the real addition result $X+Y$. When $a$ is large enough, the error term can be rounded to 0. But if the ciphertext is added infinitely, the error may increase continuously until it cannot be ignored ($>1$).

Blind multiplication is more complex. If we want to add plaintext $X$ and $Y$, we first encrypt the plaintext to get ciphertext $\left(a X+\eta_{1}\right) \bmod B$ and $\left(a Y+\eta_{2}\right) \bmod B$. The result of blind multiplication is $\left(a X+\eta_{1}\right)\left(a Y+\eta_{2}\right) \bmod B$. After decryption, we get the multiplication result of plaintext $X$ and $Y$: $\frac{\left(a X+\eta_{1}\right)\left(a Y+\eta_{2}\right)}{a^{2}}=X Y+\left(\frac{\eta_{1}}{a} Y+\frac{\eta_{2}}{a} X+\frac{\eta_{1} \eta_{2}}{a^{2}}\right)$. This result has three additional error terms $\frac{\eta_{1}}{a} Y+\frac{\eta_{2}}{a} X+\frac{\eta_{1} \eta_{2}}{a^{2}}$ than the real addition result $XY$. If ciphertext is multiplied continuously, which means that "$X$" and "$Y$" will keep increasing, the error term will keep accumulating rapidly until it will not be ignored. In addition, with the increase of times of multiplication, the error will increase faster.


\begin{table}[!t]
\caption{Amplification factor $a$ (take 4000 * 2500 = 10000000 as an example)}
\label{tab:5}
\centering
\begin{tabular}{|c|c|c|c|}
\hline
amplification factor & result & error(quantity) & error ratio(proportion)\\
\hline
10000000      & overflow &       &  \\
\hline
1000000       & 10000004 & 4     & $4 \times 10^{-6}$  \\
\hline
100000        & 10000030 & 30    & $3 \times 10^{-5}$  \\
\hline
10000         & 10000415 & 415   & $4.15 \times 10^{-4}$   \\ 
\hline
\end{tabular}
\end{table}

Due to the accumulation of errors, CMP-SWHE algorithm cannot support blind calculation of any number of times without errors. That is, CMP-SWHE is not a FHE scheme, but a SWHE scheme. We do not recommend users who require high accuracy to calculate ciphertext too many times without decryption.

An important question is, are these errors acceptable? The influence factors of error and their specific effects are quantified and analyzed as follows: We can draw a conclusion from Table~\ref{tab:5} that within a limited range, the larger the amplification factor $a$, the smaller the error. Therefore, on the premise of no overflow, we try to maximize the amplification $a$.
We can draw a conclusion from Table \ref{tab:6} that the more times of multiplication, the larger the error.
We can draw a conclusion from Table \ref{tab:7} that with the same product, the larger the difference between multipliers, the larger the error.


\begin{table}[!t]
\caption{Times of multiplication (take the actual result as 10000000 for example)}
\label{tab:6}
\centering
\begin{tabular}{|c|c|c|c|}
\hline
Polynomial (Multiplication times) & result                       & Error                    & Error ratio \\ 
\hline
10*10*10*10*10*10*10 (6) & overflow   &  &  \\
\hline
100*10*10*10*10*10 (5)                                                                     & 10003534                     & 3534                     & 0.03534\%   \\
\hline
1000*10*10*10*10 (4)                                                                       & 10002093                     & 2093                     & 0.02093\%   \\
\hline
10000*10*10*10 (3)                                                                       & 10002025                     & 2025                     & 0.02025\%   \\
\hline
100000*10*10 (2)                                                                       & 10001158                 & 1158                    & 0.01158\%   \\
\hline
1000000*10 (1)                                                                       & 10000424                     & 424                   & 0.00424\%   \\ 
\hline
\end{tabular}
\end{table}


\begin{table}[!t]
\caption{Size of multipliers  (take the real result as 10000000 for example)}
\label{tab:7}
\centering
\begin{tabular}{|c|c|c|c|}
\hline
multiplier & result   & error & Error ratio \\ 
\hline
10000*1000 & 10000006 & 6     & $6 \times 10^{-7}$           \\
\hline
20000*500  & 10000007 & 7     & $7 \times 10^{-7}$           \\
\hline
50000*200  & 10000020 & 20    & $2 \times 10^{-6}$           \\
\hline
100000*100 & 10000049 & 49    & $4.9 \times 10^{-6}$           \\
\hline
200000*50  & 10000114 & 114   & $1.14 \times 10^{-5}$           \\ 
\hline
\end{tabular}
\end{table}

In fact, times of multiplication and size of multipliers do not affect the error independently. No matter how many times the multiplication is carried out, each multiplication is carried out in turn. When one multiplication operation is finished, the result will become the multiplier of the next multiplication.

It can also be seen from Table~\ref{tab:5},~\ref{tab:6} and~\ref{tab:7} that the error and error ratio are very small in the case of limited times of calculation in ciphertext, which can be accepted in applications, especially in the fields of video and image processing that require limited and acceptable accuracy.

\subsection{Accelerated version based on batch processing}
In practical applications, it will take much time to separately perform a series of blind operations on large scale data. For example, in image and video processing, to process each pixels in a $700 \times 500$ figure, this series of operations needs to be done 350,000 times. To shorten the calculation time, we propose a batch processing version of CMP-SWHE based on the ciphertext packaging technology, which allows multiple elements to be operated at the same time. The specific operation flow of batch processing $n$ elements $\left(x_{1}, \cdots, x_{n}\right)$ is:

Take $n$ primes as a new modulo base group (or coordinate axis of the modulo) $B_{new}=\left\{B_{n e w 1}, \cdots, B_{n e w n}\right\}$ and combine these $n$ elements into a "large number" $X$ according to the Chinese remaining theorem (CRT): $X \bmod B_{n e w i}=x_{i}, i=1,2, \ldots, n $. 
It should be noted that this modulo base group $B_{n e w}$ is only owned by the client side and cannot be obtained by the server side.

After that, a series of operations in section~\ref{mod} can be performed on $X$. After the blind computation result $Y$ is obtained, we just need to do the following to obtain corresponding operation results $\left(y_{1}, \cdots, y_{n}\right)$ of each element of the original data: $Y \bmod B_{n e w i}=y_{i}, i=1,2, \ldots, n $. 
In this process, the security of our algorithm will not be reduced. On the contrary, because the server doesn't know the modulo base group $B_{n e w}$, so it doesn't know how to synthesize the large number. As a result, it makes the attacker more unable to get the original data.

In order to verify the performance of batch processing scheme, we have carried out foreground extraction processing mentioned in section~\ref{foreground} on many $576 \times 720$ images. Table~\ref{tab:11} shows the average time spent on encryption, blind calculation and decryption of each image under different levels of parallelism.

It can be seen from Table~\ref{tab:11} that the time of encryption(including time of packing ciphertext), blind calculation (blind subtraction) and decryption(including time of decomposing ciphertext) decreases in a corresponding multiple with the increase of bitch size. For example, when the batch size is 4, the encryption time is 5.014 seconds, and when the batch size is 8, the encryption time is 2.543 seconds. The former is about twice (8/4=2) as long as the latter. The reason why there is no strict multiple decrement is that, in addition to the accidental factors, there is another reason: It will take a little extra time to package the ciphertext into a batch or decompose the ciphertext of a batch. In fact, the process of package 16 elements into 1 element or decomposing 1 element into 16 elements also requires a little extra computing time. If we don't do batch processing, we don't need to spend this extra time. But this sacrifice is cost-effective, because it takes much less time to encrypt 1 element than to encrypt 16 elements.

In conclusion, experimental results show that batch processing version of CMP-SWHE greatly improves the efficiency. In addition, as the batch size increases, it takes less time. 

However, there are some limitations in this batch processing scheme: All intermediate results of all blind computing operations on each element must be smaller than its corresponding modulo base, otherwise the results obtained are not unique.


\begin{table}[!t]
\caption{Processing time required for different batch sizes}
\label{tab:11}
\centering
\begin{tabular}{|c|c|c|c|}
\hline
batch size & Encryption & Blind computing & Decryption \\ 
\hline
1 & 30.884s & 18.343s & 60.381s \\
\hline
4 & 5.014s & 4.658s & 14.498s \\
\hline
8 & 2.543s & 2.341s & 7.164s \\
\hline
12 & 1.709s & 1.547s & 4.907s \\
\hline
16 & 1.298s & 1.307s & 3.664s\\
\hline
20 & 1.042s & 0.927s & 2.912s\\ 
\hline
\end{tabular}
\end{table}

\section{CMP-SWHE Software Library }

In order to facilitate users to use this algorithm directly in various application scenarios without realizing specific algorithm details. We have integrated CMP-SWHE algorithm into a General algorithm library based on C ++ language.

\subsection{Design of the Library}
There are many issues need to be considered when integrating CMP-SWHE algorithm into a general algorithm library.

\subsubsection{Automatic Homogeneous Processing}
\label{order}
~\\
A critical issue is that the addition, subtraction, multiplication, and other operations in CMP-SWHE algorithm require the operands on both sides of the operator have the same "order". The "order" here refers to the number of times of plaintexts are encrypted. For example, in polynomial $xy + z$, $x$, $y$, and $z$ here are ciphertexts, the order on the left side of the addition operator is 2, and the order of right side is 1. At this time, according to the algorithm, the right term needs to be multiplied by the amplification $a$ to be directly added. This operation is to ensure that the real result can be restored after dividing by $a^{2}$ during decryption. But if the user doesn't know $a$, he can't make it up by himself. And even if users know it, it's too complicated for them to make up. Therefore, by storing order in the ciphertext structure, our library realizes the function of automatically complementing order before operation.

\subsubsection{Semi-blind Computation}
~\\
Sometimes users not only need to add, subtract and multiply two ciphertexts, but also need to add, subtract and multiply one ciphertext and one plaintext. Therefore, our library also supports "Semi-blind" computing, providing users with corresponding interfaces. We first encrypt the plaintext operand, and then blind calculate this ciphertext with the other ciphertext operand. If the plaintext operand comes from the server and the client is not allowed to know or unnecessary to know, the server can encrypt according to the public key instead of the private key.

\subsection{Library Performance}

\subsubsection{Algorithm performance}
\label{performance}
~\\
\begin{figure}[htbp]
\centering
\includegraphics[width=\linewidth]{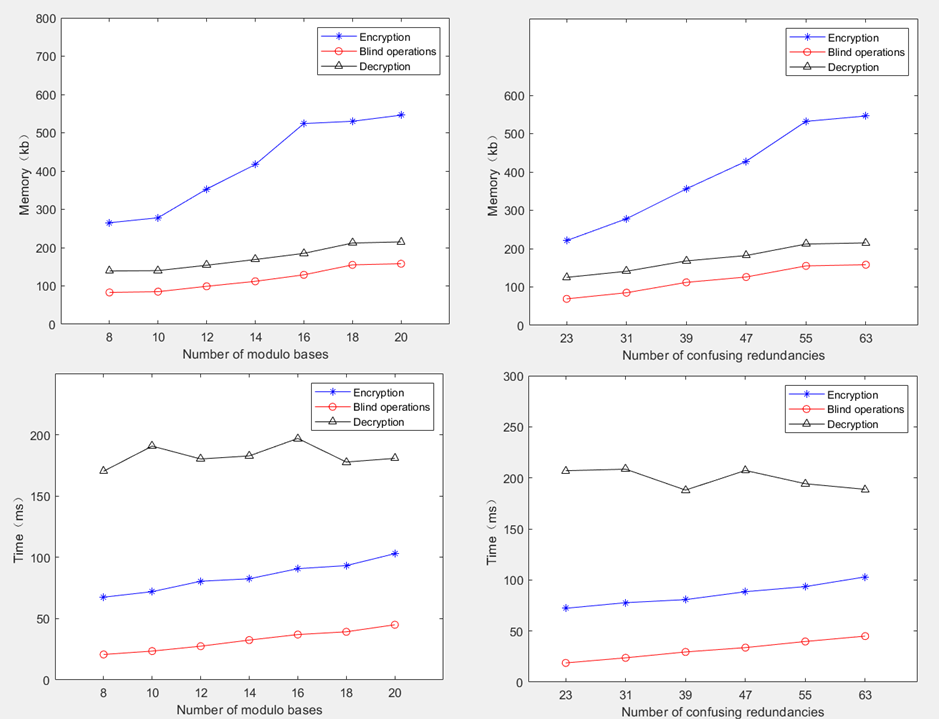}
\caption{The influence of modulo bases and confused redundancies on time and memory consumption.}
\label{fig:mod_conf}
\end{figure}

We do a series of blind calculations on $24$ sets of integers with different modulo bases and confused redundancies to test performance, including addition, subtraction, multiplication and negation. Figure~\ref{fig:mod_conf} shows the influence of confused redundancies and modulo bases on time and memory consumption. For encryption and blind calculation, more module bases and confused redundancies will undoubtedly increase time and memory consumption. On the contrary, since only the extracted correct modulo projections need to be decrypted, the effect of confused redundancies on decryption time is not in positive proportion.

In CWP-SWHE, encryption and decryption are completed on the client side, while blind operations are carried out on the server side. Taking $20$ modulo bases and $64$ confused redundancies as an example, it takes $103$ ms \& $546$ kb for encryption, $46$ ms \& $158$ kb for the series of blind operations, and $188$ ms \& $215$ kb for decryption. In other words, the computation cost of the client side and server side are $291$ ms \& $761$ kb and $46$ ms \& $158$ kb, respectively. In addition, as the client side is only responsible for data encryption and decryption, as long as the original data is the same, the computing cost of the client will not change, no matter how complex blind calculations become.

\subsubsection{Comparison with other homomorphic encryption algorithms}
We compared the performance of CMP-SWHE with four widely used fully homomorphic algorithms described in Section~\ref{sec:intro}: faster BGV scheme with bootstrapping~\cite{halevi2018faster}(implemented by IBM HElib library),a full RNS variant of BFV scheme with ~\cite{bajard2016full}(implemented by Microsoft SEAL library), CKKS scheme with bootstrapping~\cite{cheon2018bootstrapping}(implemented by HEAAN v1.1 library), improved CKKS scheme with bootstrapping~\cite{chen2019improved}(implemented by HEAAN v2.1 library). We performed the same series of operations on $24$ sets of integers, including addition, subtraction, multiplication, and negation. The experimental results with the setting of 20 modulo bases and 63 confusing redundancies in CMP-SWHE are shown in Table~\ref{tab_time_com_fhe} and~\ref{tab_mem_com_fhe}. The encryption here includes the process of parameter and key generation. Obviously, our library takes less time and memory than the other two Fully homomorphic encryption libraries. In other words, we have a huge advantage in efficiency.

\begin{table}[htbp]
\caption{Time comparison (milliseconds)}
\label{tab_time_com_fhe}
\centering
\begin{tabular}{|c|c|c|c|c|c|}
\hline
 & Ours & ~\cite{bajard2016full} & ~\cite{halevi2018faster} & ~\cite{cheon2018bootstrapping} & ~\cite{chen2019improved} \\ 
\hline
Enc. &  140 & 4909 & 25439 & 2771  & 2720 \\
\hline
Cal & 52 & 9575  & 2980  & 2022 & 1729 \\
\hline
Dec. & 243  & 1813  & 1480 & 437 & 387 \\ 
\hline
\end{tabular}
\end{table}

\begin{table}[htbp]
\caption{Memory comparison (kilobytes)}
\label{tab_mem_com_fhe}
\centering
\begin{tabular}{|c|c|c|c|c|c|}
\hline
 & Ours & ~\cite{bajard2016full} & ~\cite{halevi2018faster} & ~\cite{cheon2018bootstrapping} & ~\cite{chen2019improved} \\ 
\hline
Enc. &  553 & 13566 & 210032 &  185076 &  198520 \\
\hline
Cal & 1581 & 40071 & 219772   & 221988  &  227256 \\
\hline
Dec. & 2155 & 40702 & 219772  & 221988  &  227320 \\ 
\hline
\end{tabular}
\end{table}

We also compare CMP-SWHE with other partial homomorphism encryption and somewhat homomorphism encryption algorithms. Since most of the latest PHE and SWHE algorithms are combined with specific application scenarios, which makes it difficult for us to compare their performance under the same conditions, we choose two classical PHE and SWHE algorithms as reference: Paillier encryption scheme~\cite{paillier1999public} and BGN scheme~\cite{boneh2005evaluating}.




We encrypt, blind add and decrypt two integers several times, and average the time and memory consumption. The time and memory consumption results are shown in Table~\ref{tab:13}-~\ref{tab:14}. It can be seen that compared with the other two schemes, our scheme greatly reduces the operation time and memory consumption.

Compared with PHE schemes, our scheme has both additive homomorphic properties and multiplicative homomorphic properties. Compared with the classical SWHE scheme BGN, our additive homomorphic operation can be done many times, not only once. Compared with FHE schemes, our efficiency has a great advantage and is more suitable for application, especially for the image and video fields which does not need such high accuracy and precision. Based on these comparisons, although our scheme has errors, it is still more suitable for application. In addition, we provide users with detailed error analysis in section~\ref{errora} for reference and evaluation.

%
%
\begin{table}[htbp]
\caption{Time comparison on 3 schemes}
\label{tab:13}
\centering
\begin{tabular}{|c|c|c|c|}
\hline
 & Ours(ms) & Paillier(ms) & BGN(ms) \\ 
\hline
Encryption & 0.8 & 11 & 91 \\
\hline
Blind addition & 0.1 & 32 & 70 \\
\hline
Decryption & 0.1 & 15 & 69 \\ 
\hline
\end{tabular}
\end{table}

\begin{table}[htbp]
\caption{Memory comparison on 3 schemes}
\label{tab:14}
\centering
\begin{tabular}{|c|c|c|c|}
\hline
 & Ours(kb) & Paillier(kb) & BGN(kb) \\ 
\hline
Encryption & 9 & 1008 & 12583 \\
\hline
Blind addition & 20 & 5034 & 7340 \\
\hline
Decryption & 9 & 0 & 6814 \\ 
\hline
\end{tabular}
\end{table}


\section{Applications on Blind Vision}
\label{app}
In this section, we show how our CMP-SWHE is applied on blind vision problems, such as blind foreground extraction, blind optical flow tracking and blind object detection. These applications are very helpful for building a smart city.

\subsection{Blind Foreground Extraction}
\label{foreground}
Foreground extraction is a technology for extracting foreground objects from images or videos \cite{smoot1999background}. In reality, pixel level foreground extraction, such as extracting moving people from a surveillance video, sometimes has to be calculated on the cloud server because of the high computational overhead. But in order to protect our privacy, we don't want to expose the original video information to the server. CMP-SWHE can make the server process encrypted data without contacting the original data, which can effectively solve this problem. Common foreground extraction algorithms are: Background-Difference algorithm \cite{chien2002efficient} and Frame-Difference algorithm \cite{zhan2007improved}, Vibe algorithm \cite{barnich2010vibe}, etc.

Frame-Difference Algorithm subtract the pixels of a frame from those of its next frame to obtain a gray-scale difference image. Then, the difference image is thresholded to extract the moving region. If the frame number of two frames is $k$ and $k+1$ and their frame image is $f_{k}(x, y), f_{k+1}(x, y)$. The threshold of difference image is $T$, and the difference image is represented by $D(x, y)$. 
$
D(x, y)=\left\{\begin{array}{c}
{1,\left|f_{k+1}(x, y)-f_{k}(x, y)\right|>T} \\
{0, \text { others }}
\end{array}\right.
$. Background-Difference algorithm is similar to Frame-Difference algorithm, but it makes the difference operation between current image frame and the background image instead of the next frame. In order to avoid the influence of the change of the ambient light, the background image will be updated according to the current image frame.

\begin{algorithm}[htb]         
\caption{Blind Object Detection}             
\label{face}                  
\begin{algorithmic}[1]                
\REQUIRE ~~\\  An image                      
    
\ENSURE ~~\\  The location of the specific object in the image without leakage of the image content.                     

\STATE Assuming that there are $m$ modulo bases, the client performs corresponding modulo operations on the $m$ modulo bases respectively to obtain $m$ modulo projection subgraphs and set confusing redundancy;

\STATE Calculate the integral graph for each subgraph and pass it to the server;

\STATE For each detection window:

(1) For each weak classifier in each strong classifier, calculate the relevant haar feature value of corresponding window area of each subgraphs, and pass them to an additional server;

(2) The additional server synthesizes the feature value of the original image in this window through CRT and compares it with the threshold value of the weak classifier. Then, the additional server returns the comparison results to the computing server;

(3) The computing server adds the comparison result of each weak classifiers in the strong classifier to obtain the "strong classification figure value" and transmits it to the additional server;

(4) The additional server compares the "strong classification figure value" with the threshold of the strong classifier, and returns the result to the computing server;

(5) As long as the classification result value of a strong classifier does not reach the threshold, it is recognized that no face exists in the window. If a window passes all the strong classifiers, it can be considered that there is a face in the window.
              
\end{algorithmic} 
\end{algorithm} 


The experimental results of Foreground extraction are shown in figure \ref{fig:4}.(a) is the first frame of a surveillance video in a subway station. (b) and (c) are two adjacent video frames. (d) is the result of blind foreground extraction on (b) with Background-Difference algorithm. (e) is the result of blind foreground extraction on (b) with Frame-Difference algorithm.
%

\subsection{Blind Optical Flow Tracking}
In simple terms, optical flow is the instantaneous speed and direction of each point in the image \cite{horn1981determining}. Sometimes, we need to track the moving trail of a moving object from a surveillance video. This requires us to find positions of each moving pixels in the adjacent two frames. The general optical flow algorithm needs to find the feature points first, and then perform optical flow analysis on the feature points. We take moving points obtained in section~\ref{foreground} as feature points here. 

The classical optical flow method, such as LK optical flow method \cite{bruhn2005lucas}, assumes that the motion of two adjacent frames is relatively small. Therefore, Therefore, if we assume that the position of a motion point in the current frame is $\left(x_{1}, y_{1}\right)$, the position of this point in the previous frame $\left(x_{0}, y_{0}\right)$ is among the eight points around $\left(x_{1}, y_{1}\right)$. We call these 9 points a "window". 

By comparing the difference between the pixel values of these 8 points and the center point $\left(x_{1}, y_{1}\right)$, we can determine which point is more likely to be $\left(x_{0}, y_{0}\right)$. To reduce the accidental error, we compare the “total” pixel value of a $3 \times 3$ window centered on $\left(x_{1}, y_{1}\right)$ and other eight $3 \times 3$ windows centered on points around $\left(x_{1}, y_{1}\right)$. The "total" pixel values of a $3 \times 3$ window are compared. These operations consist of four actions: addition, subtraction, squaring, and comparison. Our algorithm can blind the first three operations. The similarity of windows needs to be compared, which requires an additional server to be performed in plaintext.


\begin{figure}[htbp]
\centering
\includegraphics[width=\linewidth]{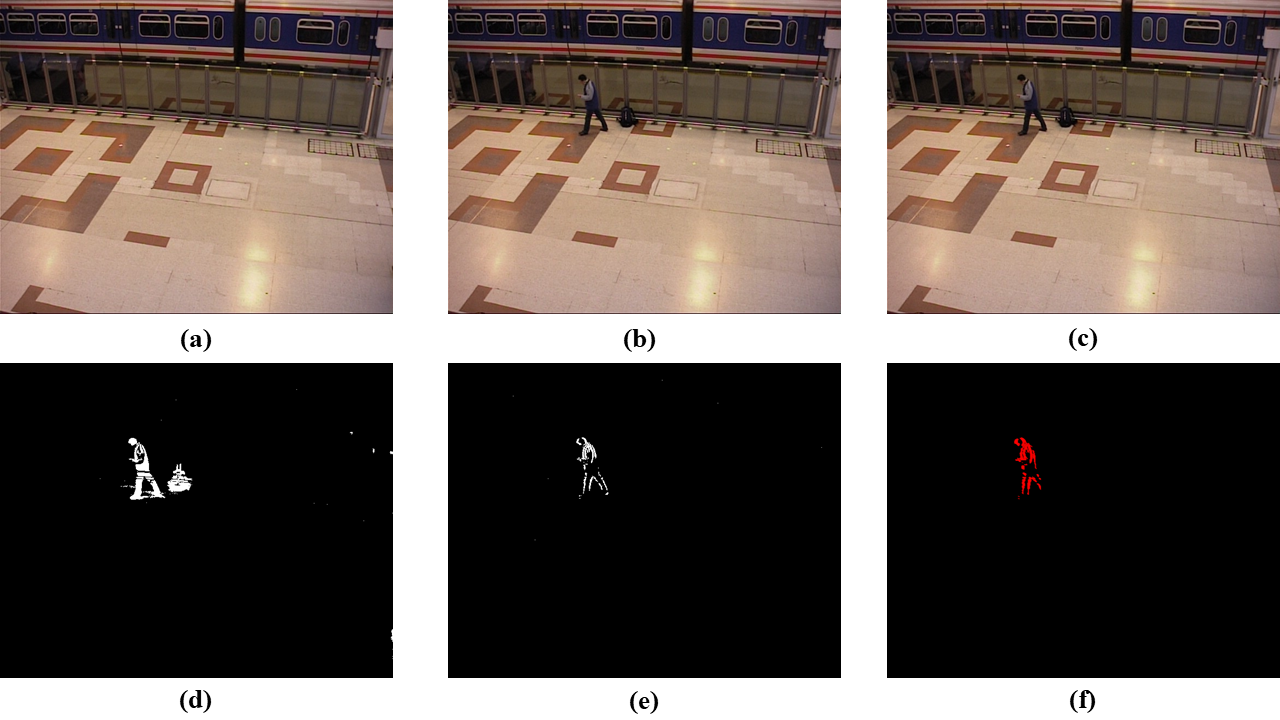}
\caption{Bind Forground extraction and Optical flow tracing results shown in plain videos}
\label{fig:4}   
\end{figure}

The experimental results of blind optical flow tracing are shown in figure~\ref{fig:4} (f). It can be seen that the noise points without motion in (e) are eliminated in (f).

\subsection{Blind Object Detection}

We implemented blind face detection by combining the Viola-Jones algorithm \cite{viola2004robust} and CMP-SWHE. The VJ algorithm introduces the concept of the integral graph to quickly calculate the haar-like features of each windows of the image. Through the AdaBoost method, multiple weak classifiers are combined into a strong classifier, which improves efficiency and ensures the accuracy of judgment. Calculating the integral graph and haar features only use the operations of addition, subtraction, and multiplication. Similarly, an additional server will be used for threshold comparison.

The specific process is shown Algorithm \ref{face}.

\begin{figure}[htbp]
  \centering
  \includegraphics[width=\linewidth]{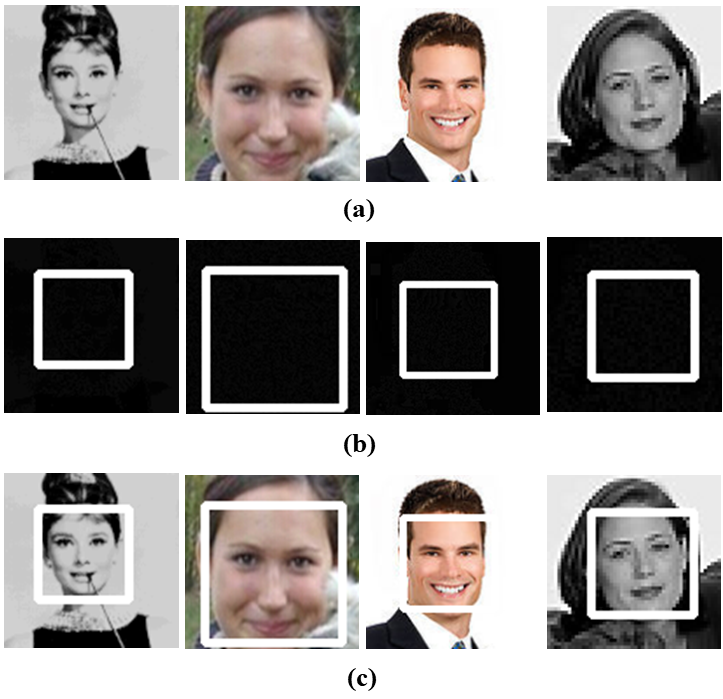}
\caption{Blind object detection results.}
\label{fig:5}       
\end{figure}

An example result of blind object detection is shown in figure \ref{fig:5}. (a) is the original image to be detected, (b) is detection result on the server side and (c) results superimposed
on original image.

An example result of blind object detection is shown in figure \ref{fig:5}. (a) is the original image to be detected, (b) is detection result on the server side and (c) results superimposed
on original image.

\section{Discussion on extending CMP-SWHE to deep learning}
After reproducing three traditional vision applications in Upmanyu's work~\cite{upmanyu2009efficient}, we try to extend CMP-SWHE to the field of deep learning. In neural network, except for the activation function, all the other calculations can be regarded as polynomial computation composed of addition, subtraction and multiplication, which can be blind processed by CMP-SWHE.

There is no doubt that handwritten digits recognition is a good entry point. In order to change it into a blind computing version, the model trained with plain MNIST dataset is encrypted on the server side and the plain test set is encrypted on the client side. Then, input the encrypted test set and encrypted model into the neural network for blind test. Finally, we can get the probability that the handwritten digit in the test image is $0-9$ after decrypting the output layer ($1 \times 10$) nodes.

However, how to deal with nonlinear activation function such as ReLU function? Polynomial approximation is a good solution. We use Chebyshev polynomial fitting method to find a polynomial by Chebyshev polynomials which is closest to the ReLU function to replace it. The recognition accuracy obtained by using the closest polynomial as the activation function should have the smallest difference with that obtained by using the ReLU function as activation function.

Another important problem is that there are many floating-point numbers in coefficients of the approximate polynomial and model, but our algorithm only supports blind calculation of integers. Therefore, it is necessary to convert all floating-point numbers in the model and activation function into integers before encryption. All floating-point numbers are uniformly rounned up to $n$ decimal places, and then multiplied by $10^{n}$. For example, "$314$" is obtained by round up "$3.1415$" to two decimal places and multiplying it by $100$.

Take a simple two-hidden-layers (100 neurons) neural network with ReLU activation function as an example. The specific process is shown Algorithm \ref{mnist}.

\begin{algorithm}[h!]         
\caption{Blind handwritten digits recognition}             
\label{mnist}                  
\begin{algorithmic}[1]                
\REQUIRE ~~\\  The test image and the model trained with plain MNIST dataset($28\times28$ images)
    
\ENSURE ~~\\   Predicted digit                    

\STATE In the original neural network, the input layer converts the test image into $layer[0]$ ($28\times28=784$), that is, the floating-point array after each pixel value in the test graph is divided by $255$. The output of layer $i$ is also the input of layer $i + 1$:
$$layer [i+1]=\operatorname{ReLU}(product[\mathrm{i}]),$$ 
\ where$$ product [i]= weights [i] \times layer [i],$$\\
\vspace{12 pt}

\STATE 
Replace $\operatorname{ReLU(x)}$ with an approximate polynomial: 
$$R(x)=k_{0} x^{m}+k_{1} x^{m-1}+\cdots+k_{m} x^{0}.$$
Then, the polynomial is processed homogeneously:
$$R_{h}(x)=k_{0} x^{m} c^{0}+k_{1} x^{m-1} c^{1}+\cdots+k_{m} x^{0} c^{m}$$
where $c=10^{n}$, $k_{i}$ are floating-point numbers.

\STATE The coefficients of $R_{h}(x)$, $weights[i]$ in the model, and $layer[0]$ are all floating-point numbers. Round up the floating-point number $f$ to $n$ decimal places and enlarge it by $10^{n}$ times to convert it into an integer:
$$f_{\text {int}}=\left\lfloor 10^{n} \times f\right\rfloor$$
\vspace{0 pt}

\STATE Encrypts the plain test data ($layer[0]$) with private key on the client side and encrypt the plain model ($weights[i]$) and $k_{i}$ in $R_{h}(x)$ with public key on the server side according to Algorithm \ref{222}.

\STATE Import encrypted test data and encrypted model into the modified neural network for testing.

\STATE If the output of each layer is magnified by the corresponding multiple, the input of the next layer will be affected. Therefore, it is necessary to modify the output of each layer as follows:
$$layer[i+1]=\frac{layer[i+1]}{\left(10^{n} \times 10^{n \times l} i\right)}=\frac{layer[i+1]}{10^{n\left(1+l_{i}\right)}},$$
where $l_{i}$ is the magnification of $layer[i]$. Since SMP-SWHE does not support division, this step is performed by additional servers.

\STATE In the output layer, the server gets an encrypted one-dimensional array ($1\times10$), that is, the probability that the handwritten digit in the test image is $0-9$, and transmits it to the client side.

\STATE Decrypt the result according to private key on client side and select index of the largest number as the predicted digit.

\end{algorithmic} 
\end{algorithm} 

However, uniformly round up all floating-point numbers to $n$ decimal places will inevitably cause some significant digits of high-precision floating-point numbers to be omitted. On the other hand, the approximate polynomial cannot be completely equivalent to the ReLU function. Therefore, we are committed to adjusting the precision of floating point to integer and improving the approximation of polynomial to reduce errors and improve recognition accuracy.

\section{Conclusions}
In this paper, we propose a single-server version of somewhat homomorphic encryption method based on the modulo projection theorem named CMP-SWHE. CMP-SWHE algorithm is based on the modulo projection theorem, the equivalence properties of congruences and the Chinese Remainder Theorem (CRT). As an extension, we also designed and implemented a batch processing version for this algorithm, which greatly improves the efficiency. 

To make this algorithm easy to use, we designed and implemented a practical general blind computing library based on CMP-SWHE. We did a lot of experiments to test the library and applied it to blind vision applications such as blind foreground extraction, blind optical flow tracking and blind face detection, which are very useful for building smart citiles. Experimental results show that although there will be errors in some cases, they are very small and acceptable, and we provide error analysis for users. In addition, our algorithm has a huge efficiency advantage, especially suitable for image and video processing and other areas with relatively low accuracy requirements. 

In the future, we will apply CMP-SWHE to more deep learning applications.

\section*{Acknowledgment}
Parts of the results and figures presented in this paper have previously appeared in our previous work~\cite{conf}. We add more technical details and experimental results in this version.

This work is partially supported by the National Natural Science Foundation of China (grant numbers 61772047), the Open Project Program of State Key Laboratory of Cryptology (grant number MMKFKT201804), the Beijing Natural Science Foundation (grant number 19L2040), the Open Project Program of State Key Laboratory of Virtual Reality Technology and Systems, Beihang University (grant number VRLAB2019C03) and the Fundamental Research Funds for the Central Universities (grant number 328201907).

%
%
%
%



\bibliographystyle{IEEEtran}
\bibliography{bare_jrnl}
\end{document}